\documentclass[]{interact}

\usepackage{epstopdf}
\usepackage{xcolor}
\usepackage[numbers,sort&compress]{natbib}
\usepackage[version=3]{mhchem}
\bibpunct[, ]{[}{]}{,}{n}{,}{,}
\renewcommand\bibfont{\fontsize{10}{12}\selectfont}
\makeatletter
\def\NAT@def@citea{\def\@citea{\NAT@separator}}
\makeatother

\begin{document}


\title{Ultrafast charge transfer and vibronic coupling in a laser-excited hybrid inorganic/organic interface}

\author{
\name{Matheus Jacobs,\textsuperscript{a} Jannis Krumland,\textsuperscript{a} Ana M. Valencia,\textsuperscript{a} Haiyuan Wang,\textsuperscript{b} Mariana Rossi,\textsuperscript{b,c} and Caterina Cocchi\textsuperscript{a}\thanks{Caterina Cocchi. Email: caterina.cocchi@physik.hu-berlin.de}}
\affil{\textsuperscript{a}Humboldt-Universität zu Berlin, Physics Department and IRIS Adlershof, D-12489 Berlin, Germany; \textsuperscript{b}Fritz Haber Institute of the Max Planck Society, Faradayweg 4-6, D-14195 Berlin, Germany; \textsuperscript{c}Max Planck Institute for Structure and Dynamics of Matter, Luruper Chaussee 149, D-22761 Hamburg, Germany }
}

\maketitle

\begin{abstract}
Hybrid interfaces formed by inorganic semiconductors and organic molecules are intriguing materials for opto-electronics. 
Interfacial charge transfer is primarily responsible for their peculiar electronic structure and optical response.
Hence, it is essential to gain insight into this fundamental process also beyond the static picture. 
\textit{Ab initio} methods based on real-time time-dependent density-functional theory coupled to the Ehrenfest molecular dynamics scheme are ideally suited for this problem.
We investigate a laser-excited hybrid inorganic/organic interface formed by the electron acceptor molecule 2,3,5,6-tetrafluoro-7,7,8,8-tetracyano-quinodimethane (F4TCNQ) physisorbed on a hydrogenated silicon cluster, and we discuss the fundamental mechanisms of charge transfer in the ultrashort time window following the impulsive excitation.
The considered interface is $p$-doped and exhibits charge transfer in the ground state.
When it is excited by a resonant laser pulse, the charge transfer across the interface is additionally increased, but contrary to previous observations in all-organic donor/acceptor complexes, it is not further promoted by vibronic coupling. 
In the considered time window of 100~fs, the molecular vibrations are coupled to the electron dynamics and enhance intramolecular charge transfer.
Our results highlight the complexity of the physics involved and demonstrate the ability of the adopted formalism to achieve a comprehensive understanding of ultrafast charge transfer in hybrid materials. 
\end{abstract}


\resizebox{25pc}{!}{\includegraphics{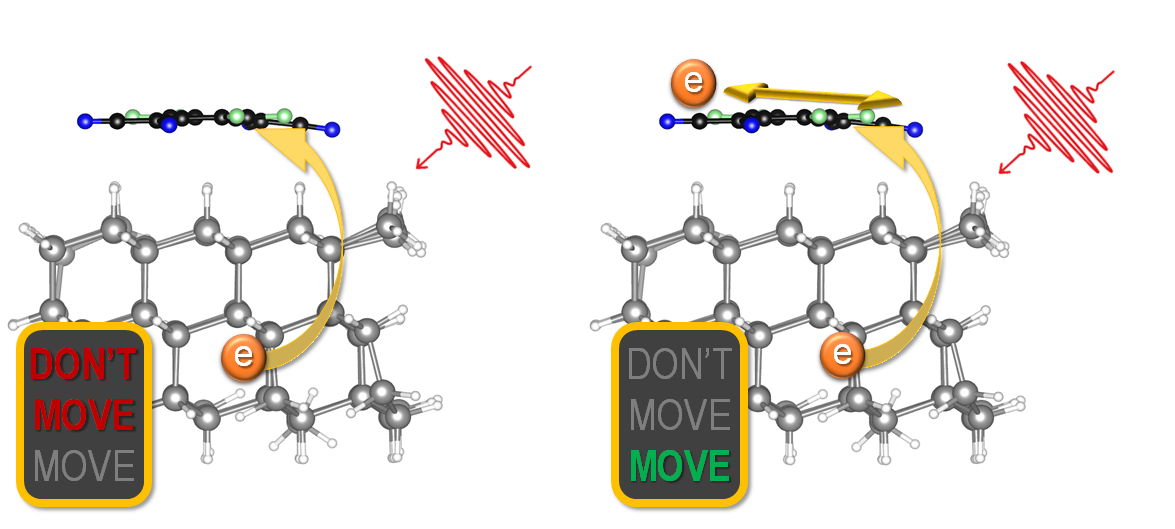}}

\section{Introduction}
Hybrid materials formed by inorganic semiconductors interfaced with organic molecules have been considered for almost three decades promising systems for novel applications in the fields of nano- and opto-electronics.
Their potential mainly resides in the unique combination of large carrier density and mobility on the inorganic side and enhanced light-harvesting and -emission characteristics of the organic components~\cite{agra+11cr,wrig12sesmc,koch12pssrrl,liu14,hewl-mcla16am,stae-rink17cp}. 
This research area was initially triggered by the successful sensitization of metal-oxide surfaces, such as \ce{TiO2} and ZnO, with molecular dyes, which allowed to obtain cost-effective solar cell devices~\cite{oreg-grae91nat}.
Since then, extensive research has been carried out on hybrid materials, demonstrating the possibility to tune the band-gap and work function of conventional inorganic semiconductors by means of molecular adsorbates~\cite{chen+09pss,hofm+13jcp,naga+13apl,feng+14jmca,corn+14pccp,schu+14afm,schu+16prb,hewl-mcla16am,schl+19prm,wang+18afm}.
In spite of the challenge that these systems and their complexity represent for \textit{ab initio} methods~\cite{drax+14acr,egge+15nl,nabo+19cm}, tremendous progress has been produced in this field by the synergistic interplay between experiments and first-principles calculations (see, e.g., Refs.~\cite{hofm+13jcp,xu+13prl,virg+13jpcc,corn+14pccp,wang2019modulation}).
More recently, the advent of two-dimensional semiconductors, such as transition-metal dichalcogenides (TMDCs), has further broadened the playground for designing hybrid interfaces.
The enhanced light-matter coupling predicted and exhibited by this emerging class of systems~\cite{mak+10prl,eda+11nl,bern+13nl,cher+14prl,li+14prb,laga+14prl,stei+17natcom,brem+18sr,mak+18natph,wang+18rmp,wen+19infomat} represents a true potential for unprecedented opto-electronic performance upon interaction with light harvesting organic compounds. 

From a fundamental viewpoint, the peculiarity of hybrid materials resides in the unique electronic properties that are generated by interfacing inorganic and organic components. 
These characteristics are determined by the interaction between delocalized bands in inorganic semiconductors and localized orbitals in organic molecules.
The electronic hybridization generates new states which do not exist in the individual components and therefore uniquely define the hybrid system~\cite{stae-rink17cp,cate-calz11jpcc,calz+12jpcc,virg+13jpcc,grue+15jpcc,kell+16am}.
The resulting level alignment between the constituents is very much dependent on the materials involved but also on the relative molecular concentration and orientation~\cite{mou+12apl,noor-gius12afm,schl+13prb,grue+15jpcc,kart+15jpcc,mile+19jpcc,chen+19jpcl,wang2019modulation}.
The introduction of a dipole layer at the interface was shown to further tune the resulting lineup~\cite{pier+15jpcl,timp+15acsami,zoje+19ami}. 
The underlying physical mechanism is the interfacial charge transfer which manifests itself already in the ground state and is primarily responsible for the flexible modulation of the electronic and optical properties of hybrid systems~\cite{mont12jpcl,xu+13prl,arna+14jpcc,matt+14aem,wu+14jpcl,grue+15jpcc,holl+17nano,hoer+18prb,meis+18jpcc,erke-hofm19jpcl,wang+19jpcc}.

Upon photo-excitation, the charge transfer and the faceted electronic structure of hybrid systems give rise to new types of excitons, characterized by a specific amount of (de)localization on the individual components and/or across the interface~\cite{gund+07pss,wu+14jpcl,eyer+15apl,fu+17pccp,pand+16prb,fogl+16cpl,ljun+17njp,eyer+17jpcc}.
A meaningful example is given by a recent first-principles study on a hybrid inorganic/organic interface formed by a monolayer of pyridine molecules chemisorbed on a ZnO surface~\cite{turk+19ats}.
In such a system, where the hybridization between the electronic states of the organic and the inorganic sides is enhanced by the covalent bond, \textit{hybrid} excitons coexist with \textit{charge-transfer} ones. 
The former are electron-hole pairs delocalized at the interface between the molecule and the inorganic surface.
Charge-transfer excitons, instead, are characterized by spatial segregation of the hole and the electron on opposite sides of the interface with only a minimal overlap.
The oscillator strength associated to charge-transfer excitons is thus significantly lower compared to hybrid excitons, in which the wave-function overlap is intrinsically larger. 
Also excitons localized solely on either component appear in the absorption spectrum of the pyridine-ZnO interface investigated in Ref.~\cite{turk+19ats}.
On account of the relative energy gaps of the inorganic and organic constituents, excitons confined on the ZnO layer appear in the low-energy region, while those localized only on the molecular monolayer are found at higher energies, in the ultraviolet band. 
A recent experimental study performed on a related interface demonstrated the predicted rationale~\cite{vemp+19jpcm}.

The identification of the nature of the excitations in hybrid interfaces represents the starting point for an in-depth understanding of the fundamental mechanisms ruling interfacial charge transfer in these materials.
Open questions concern the dynamics of charge transfer and its transient regime in the ultrashort time window immediately following the laser perturbation.
In studies performed on all-organic complexes, the coherent nuclear motion following the impulsive optical excitation was shown to be mainly responsible for the charge transfer at both covalently or non-covalently bound interfaces~\cite{rozz+13natcom,falk+14sci,pitt+15afm,desi-lien17pccp,xu19nl}.
This scenario is intrinsically different from the charge-transfer excitons mentioned above, where the electron-hole separation results from the electronic structure of the material.
It is therefore relevant to investigate the role of vibronic coupling to better understand the ultrafast charge-carrier dynamics of hybrid interfaces. 

To elucidate the mechanisms of interfacial charge transfer in photo-excited hybrid interfaces, we present a first-principles study carried out on a hybrid system formed by a hydrogenated silicon nanocluster doped by the electron accepting molecule 2,3,5,6-tetrafluoro-7,7,8,8-tetracyano-quinodimethane (F4TCNQ).
The material is excited by an ultrafast laser pulse in resonance with one of its optical excitations in the visible region. 
The interest in this structure raises from a recent joint experimental and \textit{ab initio} work investigating the change in the work function of a hydrogenated silicon surface oriented along the (111) crystallographic direction, $p$-doped by monolayers of strong molecular acceptors, including F4TCNQ~\cite{wang2019modulation}.
The resulting systems showed a remarkable degree of charge transfer in the ground state, leading to an effective tunabilty of the work function and to the partial occupation of the frontier states. 
To treat the F4TCNQ:H-Si(111) interface, we adopt a fully \textit{ab initio} non-perturbative approach based on real-time time-dependent density-functional theory (RT-TDDFT) coupled with the Ehrenfest nuclear dynamics~\cite{marques2003octopus,rozz+17jpcm}.
While model Hamiltonians have strongly contributed to rationalize charge-transfer effects also in the dynamical regime (see, e.g., Refs.~\cite{tamu+11jpcc,rens-forr14prb,desi+16natcom}), an \textit{ab initio} description of these phenomena will contribute to obtain unbiased predictions.
The laser pulse perturbing the system is explicitly included in our simulations.
In this way, we are able to access also the transient phase of the photo-excitation without any \textit{a priori} assumption or constraint on the excited state of the system.

This contribution is organized as follows: In Section~\ref{sec:theory}, we review the first-principles methodology adopted in our study and provide all the computational details.
The main body of the results is presented in Section~\ref{sec:results}, where we first describe the considered hybrid system (Subsection~\ref{sec:systems}), then we introduce its ground-state properties, including electronic structure and linear optical absorption (Subsection~\ref{ssec:gs}), and finally we analyze the laser-induced charge-carrier dynamics (Subsection~\ref{ssec:dynamics}), with particular focus on the vibronic coupling (Subsection~\ref{ssec:vibronic}).
We conclude the paper with an extended discussion of our results with respect to the current understanding of charge-transfer dynamics in donor/acceptor interfaces and the available theoretical methods to address this problem. 

\section{Theoretical Background and Computational Details}
\label{sec:theory}

\subsection{Real-time time-dependent DFT for ultrafast charge-carrier dynamics}

Density-functional theory (DFT)~\cite{hohe-kohn64pr,kohn-sham65pr} represents the most successful quantum-mechanical approach for calculating the electronic structure of materials from first principles.
Its theoretical foundation is set by the Hohenberg-Kohn theorems~\cite{hohe-kohn64pr}.
In practice, it is usually implemented through the solution of the Kohn-Sham (KS) equations~\cite{kohn-sham65pr} for an auxiliary system of independent particles, $\phi_i(\mathbf{r})$, which yield the same electron density as the interacting system: $n(\mathbf{r})=\sum_i P_i|\phi_i(\mathbf{r})|^2$, with $P_i$ being the occupation factors. 
In the time-dependent (TD) extension of DFT (TDDFT)~\cite{runge1984density}, the KS equations expressed in atomic units read
\begin{equation}
i\frac{\partial}{\partial t}\phi_i(\textbf{r},t) =\left( -\frac{\nabla_{\mathbf{r}}^2}{2}+v_{\text{KS}}[n](\textbf{r},t)\right)\phi_i(\textbf{r},t),
\label{eq:TDKS}
\end{equation}
where $\phi_i(\textbf{r},t)$ are the TD Kohn-Sham states and $v_{\text{KS}}[n](\textbf{r},t)$ the TD Kohn-Sham potential
\begin{equation}
v_{\text{KS}}[n](\textbf{r},t) = v_{\text{ions}}(\textbf{r},t) + v_{\text{ext}}(\textbf{r},t) + \int\text d^3r'\frac{n(\textbf{r}',t)}{|\textbf{r}-\textbf{r}'|} + v_{\text{xc}}[n](\textbf{r},t).
\label{eq:KSpot}
\end{equation}
The first two terms in Eq.~\eqref{eq:KSpot} describe the interaction of the electrons with the nuclei and with an external potential, respectively; the third one is the Hartree potential, and the fourth one the exchange-correlation (XC) potential. 
Since the exact form of this term is unknown, $v_{xc}$ must be approximated. 
Being nonlocal in space and time, it depends on the instantaneous density and also on the density at earlier times. 
In spite of the efforts to account for memory effects~\cite{mait+02prl,toka-park03prb,kurz-baer04jcp,wije-ulri05prl,hofm+12prl,liao+18epjb}, the adiabatic approximation is still assumed in most calculations: $v_{xc}$ is approximated using a ground-state XC potential with the TD electron density. 
In this context, the so-called \textit{adiabatic local density approximation} (ALDA)~\cite{ekar84prl}, which assumes the potential to be local in time and space, is widely used.
Regardless of its extensively discussed limitations~\cite{grim-para03chpch,cocc-drax15prb}, especially when dealing with charge-transfer excitations~\cite{dreu+03jcp,grit-baer04jcp,auts09chpch,kuri+11jctc,kumm17aem}, this approach turned out to be successful in describing the coherent charge-transfer dynamics in organic donor/acceptor complexes~\cite{rozz+13natcom,falk+14sci,pitt+15afm}.
Moreover, its reduced numerical complexity in comparison with hybrid functionals makes ALDA the scheme of choice for treating systems up to thousands atoms.
Both ALDA and hybrid TDDFT have been successfully adopted for decades to describe the optical response of molecular materials (see, \textit{e.g.}, Refs.~\cite{yaba-bert99pra,furc-ahlr02jcp,poga+02jcp,lopa-govi11jctc,li+14jctc}), in particular within the linear-response scheme proposed by Mark Casida~\cite{casi95,casi96tcc}. 

An alternative approach to access optical properties is to solve the time-dependent KS equations propagating them in real time, after the application of an instantaneous \textit{kick} ($\kappa$)~\cite{yabana1996time} in the form of a phase factor multiplying the KS wave-functions at the initial time:
\begin{equation}
\phi_i(\textbf{r},0^+) = e^{i\kappa\hat{\textbf{n}}\cdot\textbf{r}}\phi_i(\textbf{r},0^-).
\label{eq:kick}
\end{equation}
In Eq.\eqref{eq:kick} the kick $\kappa$ is a momentum and $\hat{\textbf{n}}$ indicates the propagation direction.
In this intrinsically non-perturbative scheme, the linear-response regime of photo-absorption is restored with sufficiently small values of $\kappa$ with respect to the momentum of the incident photon.
Large values of the kick are instead used to access the nonlinear response~\cite{cocc+14prl}.
The absorption spectrum is computed by Fourier transforming the induced dipole moment
\begin{equation}\label{dipole.eq}
\boldsymbol{d}(t) = -\int\text d^3r\, \textbf{r}\left[n(\textbf{r},t) - n(\textbf{r}, 0)\right],
\end{equation}
which results from the time propagation.
The achieved spectral resolution depends on the duration of the time propagation.

The advantage of RT-TDDFT over linear-response approaches is the possibility to couple straightforwardly the time-evolving KS system with an external TD electric field $\textbf{E}(t)$, which is described semi-classically in the dipole approximation within the length gauge~\cite{de2013simulating}.
In this case, the external potential in Eq.~\eqref{eq:KSpot} takes the form:
\begin{equation}
v_{\text{ext}}(\textbf{r},t) = \textbf{r}\cdot\textbf{E}(t).
\end{equation}
After the application of the TD field, one can \textit{probe} the system by applying the kick after a certain time delay~\cite{de2013simulating}.
In this way, it is possible to access transient absorption spectra, which describe the evolution of the absorption features in the explored time window.
Alternatively, one can follow the dynamics of the electron population under the effect of the applied TD field. 
The capability of RT-TDDFT to be interfaced with molecular dynamics is of paramount importance to access the effect of the electron-vibrational coupling in the ultrafast regime.
Within the Ehrenfest scheme~\cite{andrade2009modified,rozz+17jpcm}, the equation of motion for the nuclei can be solved in the classical limit~\cite{marq+12book}:
\begin{equation}\label{eq:Ehrenfest}
M_J\frac{\text d^2}{\text dt^2}\mathbf{R}_J = -\int\text d^3r\,n(\textbf{r},t)\nabla_{\textbf{R}_J}v(\textbf{r},\textbf{R}_1,\dots,\textbf{R}_M),
\end{equation}
where $M_J$ and $\textbf{R}_J$ are the mass and position of the $J^{th}$ nucleus, respectively. 
The term $v(\textbf{r},\textbf{R}_1,\dots,\textbf{R}_M)$ in the integral of Eq.~\eqref{eq:Ehrenfest} accounts for the electrostatic potential energy between the nuclei and between a nucleus and an electron in position $\textbf{r}$, and also for the coupling of the nuclei to an external field. 
The Ehrenfest formalism adopts a mean-field approach based on the calculation of the forces on the classical trajectory obtained by averaging over the quantum-mechanical degrees of freedom~\cite{tull98fd,tull12jcp,curc+13chpch,rozz+17jpcm}.
Within this approach the coupling between different vibrational modes is included and effectively participates in the interaction with the electronic excitations.
Due to its mean-field character, the Ehrenfest method is suitable to explore instantaneous processes, on the order of a few hundred fs~\cite{curc+13chpch,birc+17sd}.
Out of this regime, the Ehrenfest dynamics leads to steady-state solutions that depart from quantum detailed-balance and the corresponding results may become unphysical.

The analysis of the electron population dynamics can be performed in different ways.
When considering only the electronic motion, one can monitor the change in the population $\Delta P_j (t)$ of a certain single-particle state $\phi_j(\textbf{r})$, by projecting all the occupied time-dependent KS states $\phi_i(\textbf{r},t)$ onto $\phi_j(\textbf{r})$ and by subtracting the ground-state population of the latter:
\begin{equation}
\Delta P_j(t) = \sum_{i}^{occ}P_{i}(0)|\langle\phi_{j}(0)|\phi_{i}(t)\rangle|^2 - P_{j}(0).
\label{eq:pop}
\end{equation}
Eq~\eqref{eq:pop} is used to determine the participation of the KS states in the excitation but does not provide any quantitative information regarding the many-body excited state. 
The TD induced density, $\delta n(\textbf{r},t)$, can be computed as the difference between the electron density $n(\textbf{r},t)$ at a given time $t$ and the ground-state density at $t=0$:
\begin{equation}
\delta n(\textbf{r},t) = n(\textbf{r},t) - n(\textbf{r},0).   
\label{eq:deltan} 
\end{equation}
This quantity offers an intuitive representation of the charge localization in the laser-excited system. 
To quantify the amount of charge transfer from the donor to the acceptor in a consistent way, regardless whether the nuclear motion is explicitly taken into account or not, it is most convenient to use the Bader charge analysis~\cite{bade90}.
An alternative approach based on the Hirshfeld partition scheme was adopted in Ref.~\cite{kole+15jctc}. 

\subsection{Computational Details}

The RT-TDDFT calculations reported in this work are performed with the \textsc{octopus} code~\cite{andrade2015real, castro2006octopus, marques2003octopus}.
As a first step, we optimize the structure of the F4TCNQ:H-Si(111) interface by minimizing the interatomic forces until they are smaller than 10$^{-2}$ eV/\AA{}.
The choice of this relatively high threshold is due to the size of the system. 
The FIRE algorithm~\cite{bitz+06prl}, as implemented in \textsc{octopus}, is used for this purpose.
The KS equations are solved on a real-space grid with spacing 0.16~\AA{} within a simulation box formed by interlocked spheres of radius 10.5~\AA{} centered at each atomic position.
A Gaussian smearing of 0.001~eV is included to ensure a smooth convergence of the KS equations (Eq.~\ref{eq:TDKS}) in the presence of partial occupations of the electronic states.
Core electrons are treated using Hartwigsen-Goedecker-Hutter norm-conserving pseudopotentials~\cite{hartwigsen1998relativistic, goedecker1996separable}. 
The local-density approximation (LDA) in the Perdew-Zunger parameterization~\cite{perd-zung81prb} is adopted to approximate the XC potential.
Due to the well-known error cancellation embedded in this functional, it is not necessary to add an additional correction to account for the van der Waals forces~\cite{maro+11jctc}, which dominate the interaction between the donor and the acceptor in the considered hybrid interface.
Optical properties in the static and dynamical regime are computed by propagating in time the KS equations (Eq.~\ref{eq:TDKS}) using the approximated enforced time-reversal symmetry propagator \cite{castro2004propagators} with a time step of 1.21 attosecond and a total propagation duration of approximately 40 fs.
The optical spectra are computed according to the Yabana-Bertsch scheme~\cite{yabana1996time}, applying a kick $\kappa =$~0.0053~\AA{}$^{-1}$ (see Eq.~\ref{eq:kick}) along all three Cartesian directions.
To study the laser-induced dynamics, a resonant TD electric field with carrier frequency $\hbar \omega = 2.18$~eV (corresponding to a wavelength of 580 nm) and total duration 20 fs is applied. 
The pulse has trapezoidal shape with ascending and descending ramps of duration 5~fs each, and constant amplitude of 10~fs corresponding to an intensity of 500~GW/cm$^{2}$.
This value is chosen to ensure an effective perturbation of the system without causing ionization. 
The Ehrenfest scheme is applied assuming the system to be at 0~K, in order to rule out thermal excitations from the analysis of the coupled electron-vibrational dynamics. 

For the Bader charge analysis the time-dependent electron density is calculated at intervals of 1.5 fs.
In the Bader charge analysis and in the single-particle population analysis the Savitzky-Golay filter~\cite{savitzky1964smoothing,press1992numerical} is applied to smooth the oscillations in the total density.
This approach consists in a convolution method that increases the precision of the data without loosing the tendency. 
For our purposes, we use a 7$^{th}$-order polynomial with 81 window points.

\textit{Ab initio} molecular dynamics simulations for F4TCNQ in the gas phase and adsorbed on the H-Si(111) cluster are performed with the i-PI program~\cite{Kapil2019} in connection with the FHI-aims code~\cite{blum+09cpc}.
The LDA for the XC functional in the Perdew-Wang parameterization~\cite{perd-wang92prb} is adopted along with standard \textit{light} settings for all atomic species in the FHI-aims code. 
We use a 1~fs time step and simulate the system in the canonical ensemble by coupling it to the stochastic velocity rescaling thermostat~\cite{Bussi2007}. 
We run 5~ps of thermalization at 300~K and subsequently gather statistics from 20~ps of simulations from two trajectories for the F4TCNQ:H-Si(111) hybrid system and from 50~ps of simulations for the free molecule.

\section{Results and discussion}
\label{sec:results}
\subsection{The F4TCNQ:H-Si(111) hybrid interface}
\label{sec:systems}

\begin{figure}
    \centering
    \includegraphics[width=1.0\linewidth]{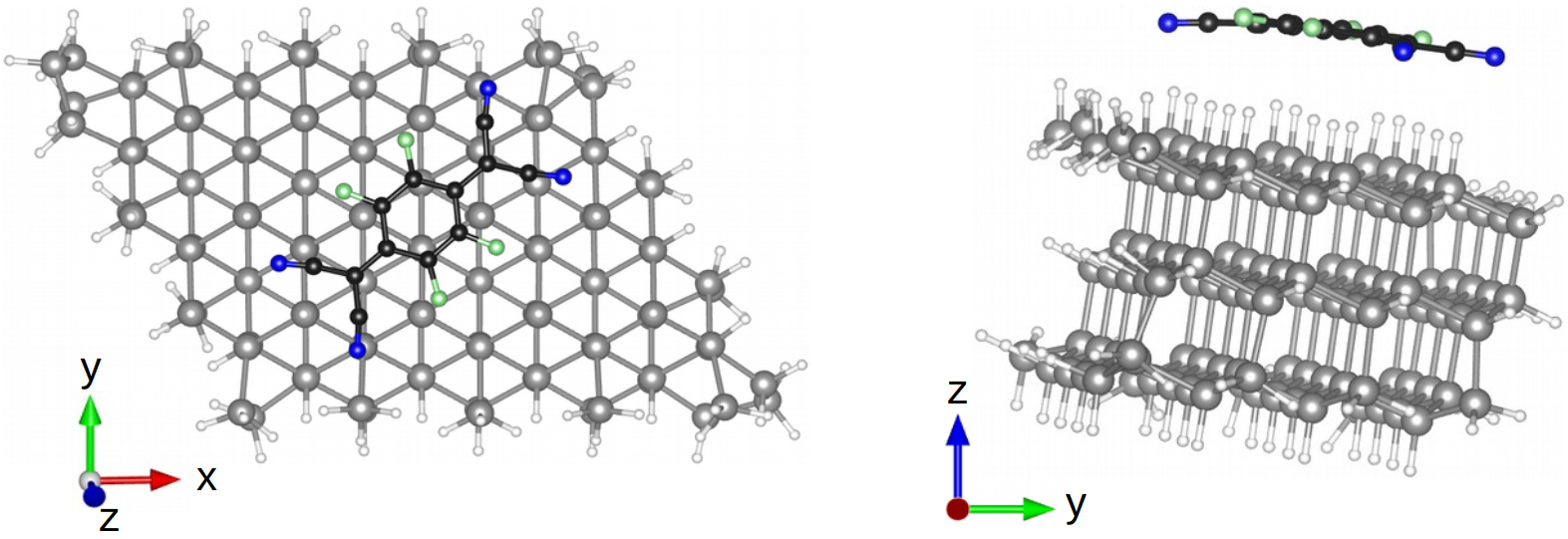}
    \caption{(left) Top and (right) side view of the F4TCNQ:H-Si(111) hybrid interface cluster. Carbon atoms are depicted in black, nitrogen in blue, fluorine in green, silicon in grey, and hydrogen in white.}
    \label{fig:system}
\end{figure}

The interface formed by the H-Si(111) surface and the adsorbed electron-accepting molecule F4TCNQ was previously studied in Ref.~\cite{wang2019modulation}.
In that case, the inorganic substrate was modeled by a semi-infinite slab including 5 double layers of Si atoms passivated with H atoms in the non-periodic direction.
Supercells of different sizes including up to 6 Si(111) unit cells in the in-plane periodic directions were considered, in order to search for the optimal structure doped by an increasing number of molecular layers, ranging from one to four. 
In the present study, we are not interested in addressing such vast structural complexity, but, conversely, in identifying a simplified and yet physically meaningful system for the study of the ultrafast laser-induced charge-transfer dynamics.
For this purpose, we model the interface with a hydrogenated Si cluster, composed of 3 atomic double layers oriented along the (111) crystallographic direction and interacting with one F4TCNQ molecule physisorbed on its surface (see Fig.~\ref{fig:system}). 
In the planar direction with respect to the molecule, the Si cluster is represented by a $(4 \times 3)$ supercell. 
The resulting system, containing 248 atoms in total, bears similarities with the hybrid interface formed by a hydrogenated diamond nanocluster doped by the perylenetetracarboxylic-acid-diimide molecule, which was previously considered to study ultrafast charge transfer with an atomistic computational approach~\cite{medr-sanc18jpcl}.

After the structural optimization (see details in Section~\ref{sec:theory}), we obtain the geometry shown in Fig.~\ref{fig:system}, with F4TCNQ at a distance of about 4.3~\AA{} from the surface.
As visible on the right panel of Fig.~\ref{fig:system}, F4TCNQ is slightly bent, forming an angle of 157$^{\circ}$ with respect to the planar configuration, similar to the case in which the same molecule is adsorbed on ZnO~\cite{xu+13prl}, graphene~\cite{kuma+17acsn}, and various hydrogenated silicon surfaces~\cite{wang2019modulation,wang+17jpcc,carv+11prb}.  

\begin{figure}[h!]
    \centering
    \includegraphics[width=\textwidth]{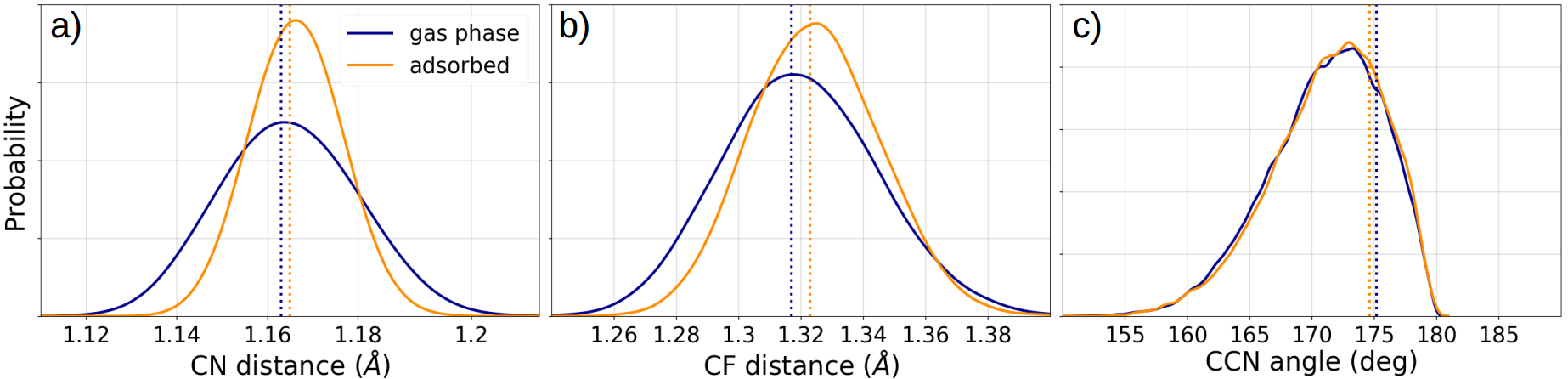}
    \caption{Normalized probability distribution obtained from \textit{ab initio} molecular dynamics at 300~K of a) C$\equiv$N and b) C--F bond distances as well as of c) CCN bond angles in gas-phase F4TCNQ and in F4TCNQ adsorbed on the H-Si(111) cluster. Dashed lines represent the optimized-structure values of each quantity.}
    \label{fig:MD1}
\end{figure}

In order to gain information about thermal effects, we investigate the influence of nuclear fluctuations at room temperature on geometrical aspects of the structure. 
From the \textit{ab initio} molecular dynamics simulations we observe that the average C$\equiv$N and C--F bond distances do not change appreciably from the respective relaxed equilibrium structure values, as shown in Fig.~\ref{fig:MD1}. 
However, their distribution is narrowed upon adsorption on the surface. 
Especially in the case of the C$\equiv$N stretch, we ascribe this behavior to the larger amount of charge accumulated on the N atoms in the adsorbed molecule compared to the gas-phase one (see also Ref.~\cite{zhu+11cm}), as discussed in Section~\ref{ssec:gs} below.
The average CCN angles are reduced by about 4$^{\circ}$ at 300~K if compared to the optimized structures, which reflects the anharmonicity of the related bending mode. 
We have checked that this deformation does not change the electronic orbitals of the molecule. 
From this analysis, we expect that nuclear fluctuations up to room temperature will not induce qualitative changes on the electron dynamics following the electronic excitation, when compared to simulations that start from optimized structures at the potential energy surface.

\subsection{Electronic structure and charge transfer in the ground-state}
\label{ssec:gs}

\begin{figure}
    \centering
    \includegraphics[width=0.8\linewidth]{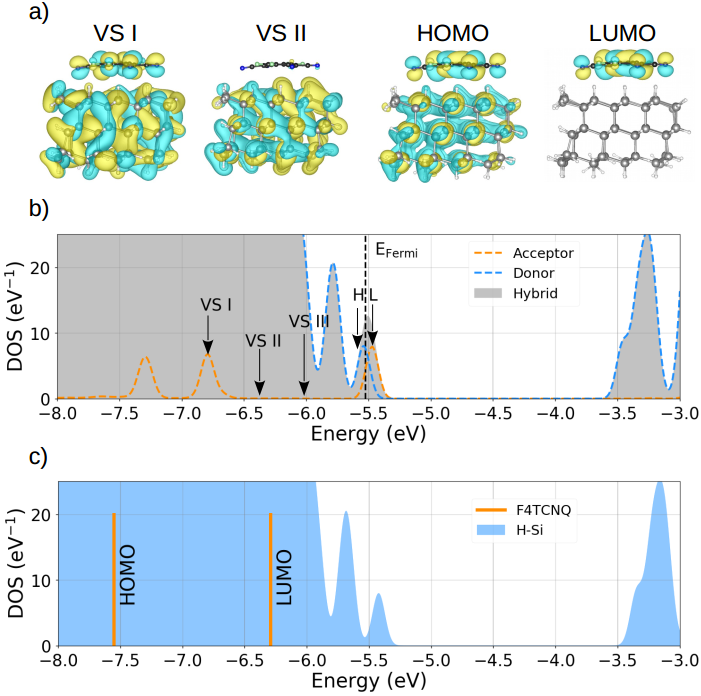}
    \caption{a) Isosurfaces of selected Kohn-Sham states of the F4TCNQ:H-Si(111) hybrid interface; b) Density of states (DOS, grey area) with projected contributions from the donor (blue dashed line) and the acceptor (orange dashed line) side of the interface. HOMO (H) and LUMO (L) as well as two valence states, VS I and VS II, plotted in panel a) are highlighted; c) Level alignment of the building blocks shown by the DOS of the isolated H-Si(111) cluster (blue area) with the frontier orbitals of the gas-phase F4TCNQ marked by orange bars. A Gaussian broadening of 0.05~eV is applied to the DOS in panels b) and c).}
    \label{fig:GS}
\end{figure}

We start our analysis of the ground-state properties of the prototypical F4TCNQ:H-Si(111) hybrid interface by examining the electronic structure of this system (see Fig.~\ref{fig:GS}) in terms of the spatial distribution of its orbitals (panel a) and of the projected density of states (PDOS -- panel b).
In Fig.~\ref{fig:GS}c), the density of states of the isolated inorganic and organic components are displayed to show their mutual level alignment.
According to the conventional nomenclature of semiconductor heterojunctions~\cite{ihn10book}, this is a \textit{type-III} interface with the frontier orbitals of the H-Si(111) cluster being both at higher energy than those of F4TCNQ.
This result is qualitatively different from the \textit{type-II} level alignment resulting in all-organic interfaces when, for example, F4TCNQ acts as a dopant of thiophene oligomers~\cite{zhu+11cm,mend+15natcom,vale-cocc19jpcc,vale+20pccp}.
Due to this level alignment, when the electron-accepting molecule and the H-Si(111) cluster interact with each other forming the hybrid interface, the latter is $p$-doped, with both the highest-occupied molecular orbital (HOMO) and the lowest-unoccupied molecular orbital (LUMO) being partially occupied. 
The HOMO hosts a charge of 1.66 $e$ while in the LUMO the fraction of electron amounts to 0.34 $e$.
We note in passing that in the literature (see, \textit{e.g.}, Ref~\cite{erke-hofm19jpcl}) these frontier states are also indicated as SOMO and SUMO, standing for \textit{singly-occupied} and \textit{singly-unoccupied} molecular orbital, respectively. 
After a close inspection of Fig.~\ref{fig:GS}a), it is evident that both the HOMO and the LUMO of the hybrid interface receive contributions from the LUMO of F4TCNQ.
However, they exhibit substantially different nature: The HOMO is delocalized across the interface and results from the hybridization between the HOMO of the Si cluster and the LUMO of the molecular acceptor.
In contrast, the LUMO of the hybrid corresponds almost identically to the LUMO of the isolated molecule (see, e.g., Refs.~\cite{zhu+11cm,vale-cocc19jpcc}), with only very small contributions coming from the hybridization with the inorganic part of the interface.
A gap of approximately 2 eV separates the LUMO to the higher states in the conduction band, which are localized solely on the H-Si(111) cluster in the energy window shown for the PDOS (see Fig.~\ref{fig:GS}b).
Also the valence region below the HOMO is dominated by Si-like states.
One of them, marked as VS II, is visualized in Fig.~\ref{fig:GS}a).
The other one, labeled VS III, has analogous nature and spatial distribution.
The contribution from molecular states in the valence region displayed in Fig.~\ref{fig:GS}b) is given by the two peaks at -6.82~eV and -7.28~eV, respectively.
The former, labeled VS I, is formed by the HOMO of F4TCNQ hybridized with an occupied Si state (see Fig.~\ref{fig:GS}a).
The latter (not shown) bears contribution from the HOMO-1 of the molecule. 
By comparing Fig.~\ref{fig:GS}b) with Fig.~\ref{fig:GS}c), we notice the energetic up-shift of the HOMO and the LUMO of F4TCNQ, by about 0.8~eV and 0.9~eV, respectively, due to the interaction of the molecule with the inorganic substrate.
In contrast, the DOS of the Si nanocluster is not significantly affected by the presence of the adsorbed F4TCNQ. 

The electronic properties of the F4TCNQ:H-Si(111) hybrid interface are consistent with the (partial) charge transfer from the inorganic donor to the organic acceptor, previously discussed in analogous systems~\cite{carv+11prb,wang+17jpcc,wang2019modulation}. 
Based on the Bader charge analysis~\cite{bade90}, the charge transfer in the ground state amounts to 0.37 $e$ in the investigated system.
This value is of the same order as the one obtained in Ref.~\cite{wang2019modulation} using density difference partition in real space for periodic F4TCNQ:H-Si(111) interfaces investigated from DFT with a hybrid XC functional~\cite{kruk+06jcp}.
The charge transferred from the H-Si(111) cluster to the molecule is distributed mainly among the N atoms (about 0.32~$e$ overall) and for the remaining among the C atoms forming the single bonds adjacent to the C$\equiv$N ones. 
The partial charges on the F atoms in the physisorbed molecule are essentially unaltered compared to gas-phase F4TCNQ. 

\begin{figure}[h!]
    \centering
    \includegraphics[width=0.9\textwidth]{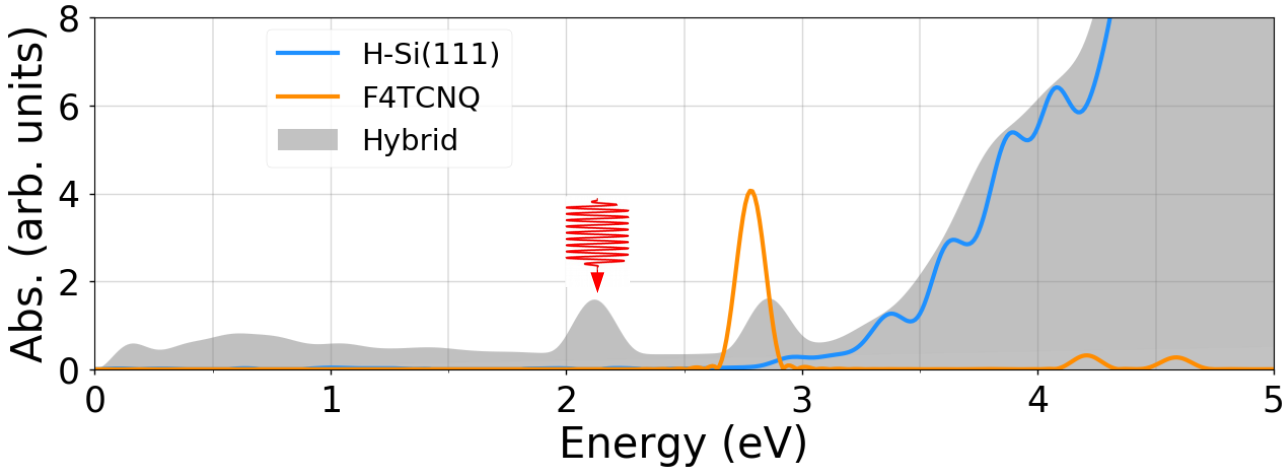}
    \caption{Linear absorption spectrum of the F4TCNQ:H-Si(111) hybrid interface (grey area), of the gas-phase F4TCNQ (orange line), and of the H-Si(111) cluster (blue line). The hybrid excitation at 2.18 eV in resonance with the applied pulse frequency is highlighted by the arrow.}
    \label{fig:opt}
\end{figure}

The optical spectrum of F4TCNQ:H-Si(111) computed from RT-TDDFT is shown in Fig.~\ref{fig:opt}.
As an effect of $p$-doping (see Fig.~\ref{fig:GS}b), non-vanishing absorption appears starting from very low energies in the infrared region up to approximately 2 eV.
Allowed transitions to the LUMO from the HOMO and from lower electronic states in its vicinity contribute to the weak absorption in this energy range. 
At 2.18 eV a pronounced peak is visible, followed by another one at 2.89 eV of very similar oscillator strength. 
A steep absorption onset appears starting from about 3 eV.
To interpret these features, we inspect the spectra of the isolated molecule (orange line) and of the hydrogenated silicon cluster (blue line) overlaid to that of the hybrid interface in Fig.~\ref{fig:opt}.
From this comparison, it is evident that the sizable absorption of F4TCNQ:H-Si(111) above 3 eV stems from electronic transitions within the Si cluster.
On the other hand, the peak at 2.89 eV almost coincides in energy with the lowest-energy excitation in the gas-phase F4TCNQ.
This similarity does not exclude contributions to this peak also from the hybridized states in the F4TCNQ:H-Si(111) interface, according to the PDOS shown in Fig.~\ref{fig:GS}b).
However, the molecular signature of this peak is predominant. 
In contrast, the peak at 2.18 eV does not have any apparent counterpart in the spectra of the individual components and can therefore be regarded as a new characteristic of the hybrid system. 
For this reason, we consider in the following the ultrafast dynamics induced by a laser pulse in resonance with this excitation at 2.18 eV.

\subsection{Laser-induced electron dynamics}
\label{ssec:dynamics}

\begin{figure}[h!]
    \centering
    \includegraphics[width=0.9\linewidth]{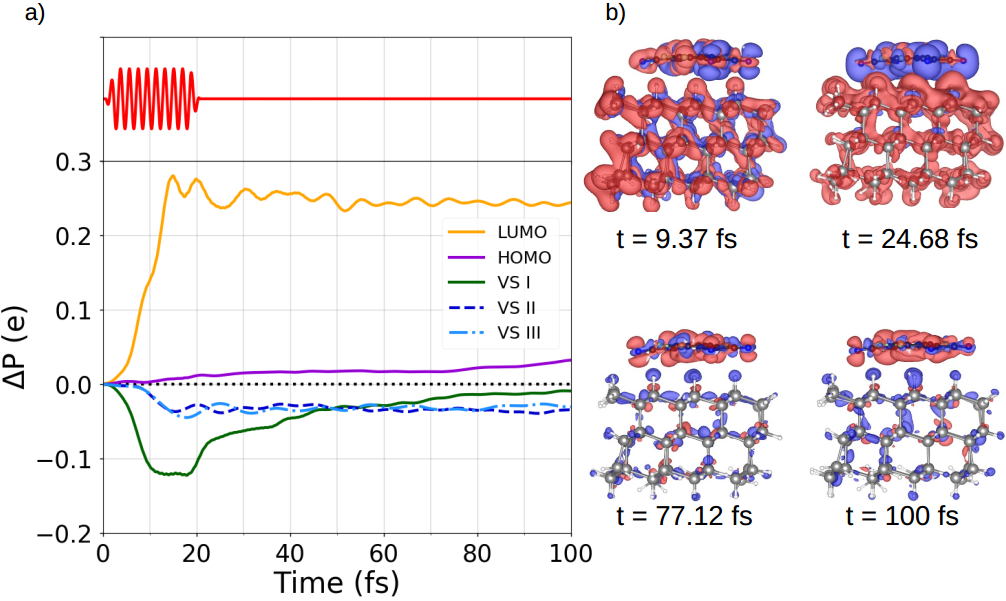}
    \caption{a) Population dynamics of the Kohn-Sham states in response to the laser pulse shown on the top panel, computed as a variation of the occupation factors $P_j$ with respect to their ground-state values; b) Snapshots of the induced density at different instants in time. Red (blue) domains of the isosurfaces indicate electron accumulation (depletion). The isosurfaces are depicted with a cutoff value of 0.00015~$bohr^{-3}$.}
    \label{fig:pop}
\end{figure}

To inspect the laser-induced dynamics of the F4TCNQ:H-Si(111) hybrid interface, we excite the system with an ultrashort pulse in resonance with the hybrid excitation at 2.18 eV (see Fig.~\ref{fig:GS}).
We represent the laser with a pulse of trapezoidal shape and total duration 20 fs.
As detailed in Section~\ref{sec:theory}, the pulse has constant amplitude for 10 fs and ascending and descending ramps of 5 fs each (see top panel of Fig.~\ref{fig:pop}a).
The pulse is polarized in the $x$ direction (see coordinate system in Fig.~\ref{fig:system}), in resonance with the polarization of the excitation at 2.18 eV in the spectrum of the interface (see Fig.~\ref{fig:opt}).

First, we focus on the laser-induced electronic dynamics, neglecting the coupling to the nuclear motion, and analyze the time-evolution of the single-particle state population, computed according to Eq.~\eqref{eq:pop} (see Fig.~\ref{fig:pop}a).
In particular, we follow the change in the occupation with respect to the ground state of the semi-occupied frontier states HOMO and LUMO, as well as of the occupied orbitals VS I, VS II, and VS III, marked in Fig.~\ref{fig:GS}b).
In the time window in which the laser is active, the LUMO population undergoes a steep increase by 0.3~$e$, which is mirrored by a decrease on the order of 0.1~$e$ in the occupation of the hybrid orbital VS I.
In the first 20 fs, corresponding to the time window of the laser excitation (see Fig.~\ref{fig:pop}), we notice a non-negligible change in the occupation of VS II and VS III, which are localized on the silicon cluster only (see Fig~\ref{fig:GS}a).
This result represents an indirect indication that the pumped excitation corresponds to a transition from the occupied VS I to the LUMO. 
After 20 fs, when the pulse is turned off, the LUMO population is almost saturated to an increase of 0.25~$e$ with respect to the ground state.
Spurious oscillations that remain after the application of the filter function (see details in Section~\ref{sec:theory}) are ascribed to the evolution of the Hartree and XC potential in the time-dependent KS equations (see Eq.~\ref{eq:TDKS}).
Also the two occupied states VS II and VS III are similarly saturated after about 20~fs.
Conversely, the hybridized orbital VS I regains almost entirely its ground-state occupation within 100 fs. 
After 20~fs also the partially occupied HOMO starts to increment its occupation up to 0.05~$e$ with respect to the ground state at the end of the simulation window. 
These results indicate a much more complex electron dynamics compared to the picture obtained in linear response.

To gain additional insight into these dynamical processes, it is relevant to inspect the temporal behavior of the induced electron density $\delta n (\mathbf{r},t)$ (see Eq.~\ref{eq:deltan}), displayed in Fig.~\ref{fig:pop}b).
After 10 fs from the beginning of the simulation, when the pulse has reached its maximum amplitude, $\delta n (\mathbf{r},t)$ is delocalized across the whole system and oscillates along the polarization direction of the laser pulse (see Fig.~\ref{fig:pop}a).
After about 25 fs, shortly after the laser has been switched off, interfacial charge transfer occurs, as testified by the electron depletion and accumulation domains visualized in Fig.~\ref{fig:pop}b).
In the last part of the simulation, from 77~fs to 100~fs, the induced charge density tends to be more localized on the molecule compared to the H-Si(111) nanocluster.
On either sides of the interface, both electron depletion and accumulation domains are visible, although the former (latter) are predominant on the inorganic (organic) side.
The electron depletion in the inorganic component is particularly localized around the H atoms that passivate the Si(111) surface facing the adsorbed molecule. 

\begin{figure}[h!]
    \centering
    \includegraphics[width=0.95\linewidth]{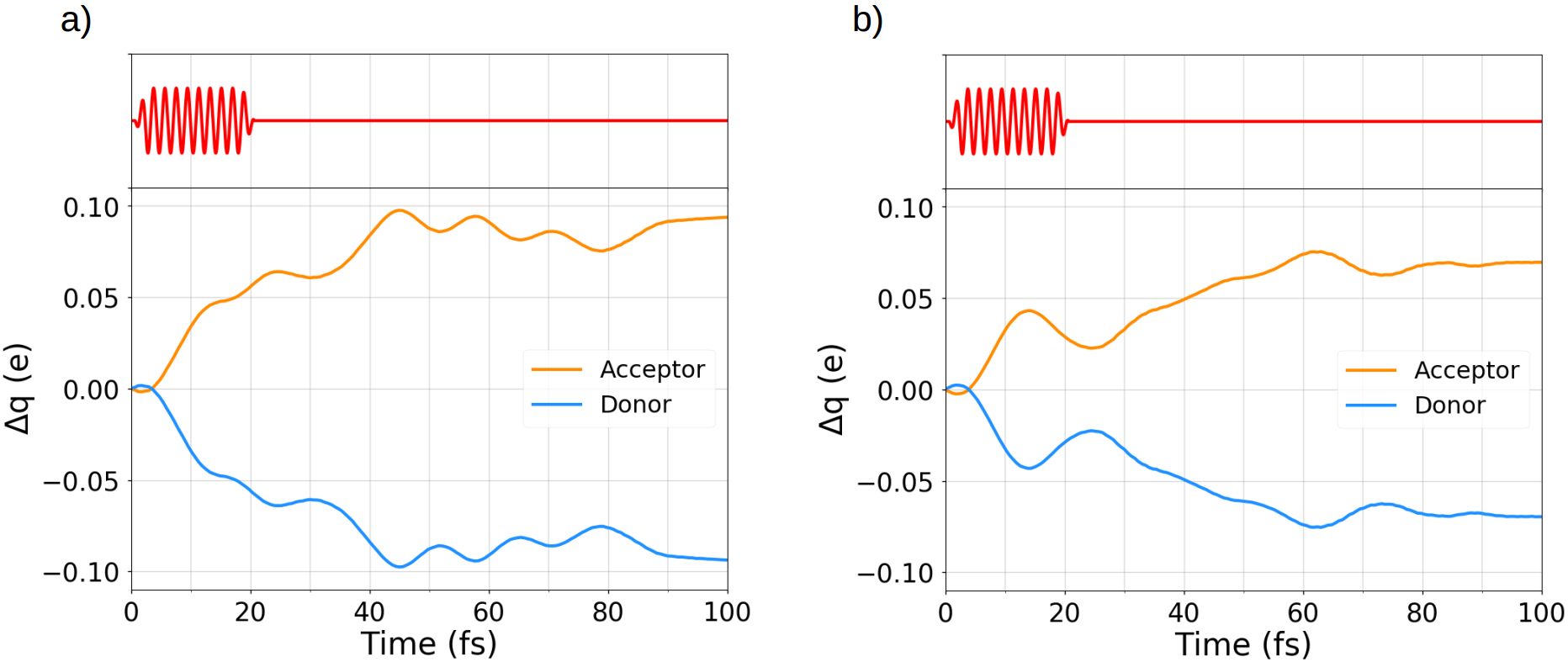}
    \caption{Time evolution of the Bader charge distribution with respect to the ground state, computed a) without and b) with vibronic coupling included.}
    \label{fig:bader}
\end{figure}

The picture provided by the dynamics of the induced electron density points to a laser-induced interfacial charge transfer from the silicon nanocluster to the molecule. 
To analyze this behavior quantitatively, we inspect the time evolution of the Bader charge distribution upon partitioning the F4TCNQ:H-Si(111) hybrid interface into two regions corresponding to the inorganic donor and the organic acceptor.
The result displayed in Fig.~\ref{fig:bader}a) confirms the trend predicted by the analysis of the occupations of the single-particle states.
The electron depletion in the donor, of the order of 0.1 $e$, is mirrored by the electron accumulation in the acceptor, with the maximum achieved immediately after the laser is turned off.
This finding implies that, upon the action of the resonant pulse, the interfacial charge transfer is enhanced by about one tenth of an electron with respect to the ground state. 
To assess the role of vibronic coupling in the charge-transfer process, we analogously analyze the TD Bader charges also when the nuclear motion is enabled in the Ehrenfest scheme.
As shown in Fig.~\ref{fig:bader}b), the charge transfer from the inorganic donor to the organic acceptor takes place in a similar fashion as with clamped ions.
However, the electron-nuclear coupling does not significantly enhance the charge transfer, as instead observed and predicted in all-organic donor/acceptor interfaces~\cite{rozz+13natcom,falk+14sci,pitt+15afm}.
The interpretation of this result demands additional analysis on the vibrational dynamics.

\subsection{Interfacial or intramolecular charge transfer? The role of vibronic coupling}
\label{ssec:vibronic}

\begin{figure}
    \centering
    \includegraphics[width=0.95\textwidth]{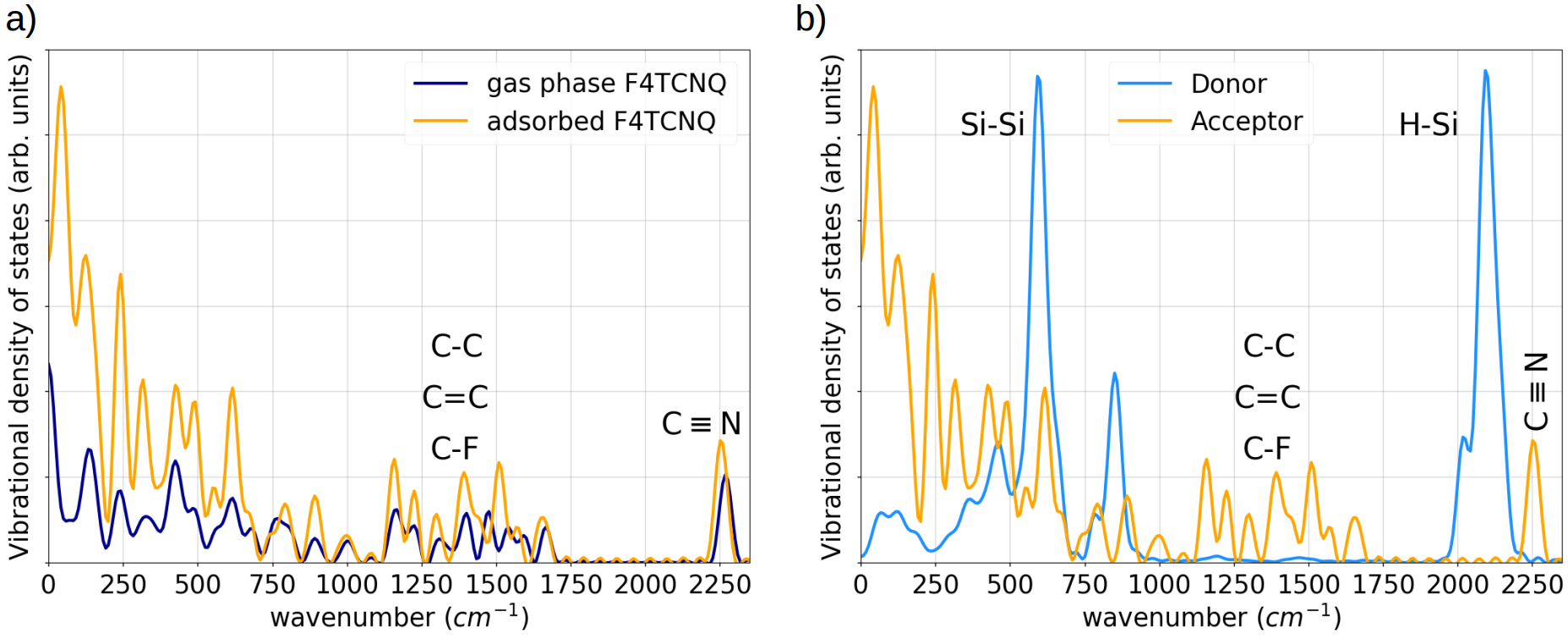}
    \caption{Vibrational spectra computed by Fourier transforming the velocity autocorrelation functions obtained from the molecular dynamics simulations of a) F4TCNQ in the gas phase (dark blue) and absorbed on H-Si(111) (orange) and b) of the projected contributions on the donor H-Si(111) (blue) and on the acceptor F4TCNQ (orange) of the hybrid interface. The contributions of the donor are divided by 6 to enhance the visibility of those of the acceptor. The labels mark the peaks associated to selected stretches. In the region between 1200~cm$^{-1}$ and 1600~cm$^{-1}$ the contributions of the C-C, C=C, and C-F modes overlap.}
    \label{fig:MD2}
\end{figure}

In order to understand the behavior of charge transfer in the F4TCNQ:H-Si(111) hybrid interface, we analyze in details the vibrational modes that participate in this process and their dynamics. 
As a first step, we inspect the vibrational spectra obtained from the Fourier transform of velocity autocorrelation functions from the \textit{ab initio} molecular dynamics simulations performed with i-PI~\cite{Kapil2019} and FHI-aims~\cite{blum+09cpc}.
In Fig.~\ref{fig:MD2}a) the results computed for F4TCNQ in the gas phase and adsorbed on H-Si(111) show that the characteristic frequencies of the molecule are overall the same in both configurations. 
We note that the spectrum of the gas-phase molecule includes molecular rotations, which explains the peak structure and the increasing background at low frequencies. 
In the hybrid F4TCNQ:H-Si(111) cluster, they become pronounced hindered rotations which present a similar, but comparatively more intense structure.
Shifts of some modes are visible in Fig.~\ref{fig:MD2}a).
The most relevant one concerns the C$\equiv$N stretch, which is usually analyzed to estimate the degree of charge transfer in the ground state of donor/acceptor complexes formed by molecules hosting this bond (see Refs.~\cite{chap+81jacs,maen+01prb,stir+05cc,zhu+11cm,beye+19cm}).
Indeed, from Fig.~\ref{fig:MD2}a) we notice that, when F4TCNQ is adsorbed on the H-Si(111) cluster, the C$\equiv$N peak is red shifted (\textit{i.e.}, softened) compared to the gas phase, as a signature of bonding and charge transfer~\cite{zhu+11cm}.
The comparison between the projected contributions of F4TCNQ and H-Si(111) in the vibrational spectra of the F4TCNQ:H-Si(111) interface (see Fig.~\ref{fig:MD2}b) reveals the relative energies of the characteristic modes of the hybrid system. 
The low-energy region of the spectrum is dominated by the peak of the Si-Si modes at about 600~cm$^{-1}$.
At higher energies, in the vibrational gap of the H-Si(111) cluster, the stretches of the C-F bonds as well as of the single and double carbon bonds appear. 
Finally, the frequency range between 2000~cm$^{-1}$ and 2500~cm$^{-1}$ is characterized by the C$\equiv$N and H-Si stretches, which, however, exhibit negligible mutual overlap.

\begin{figure}[h!]
    \centering
    \includegraphics[width=0.9\linewidth]{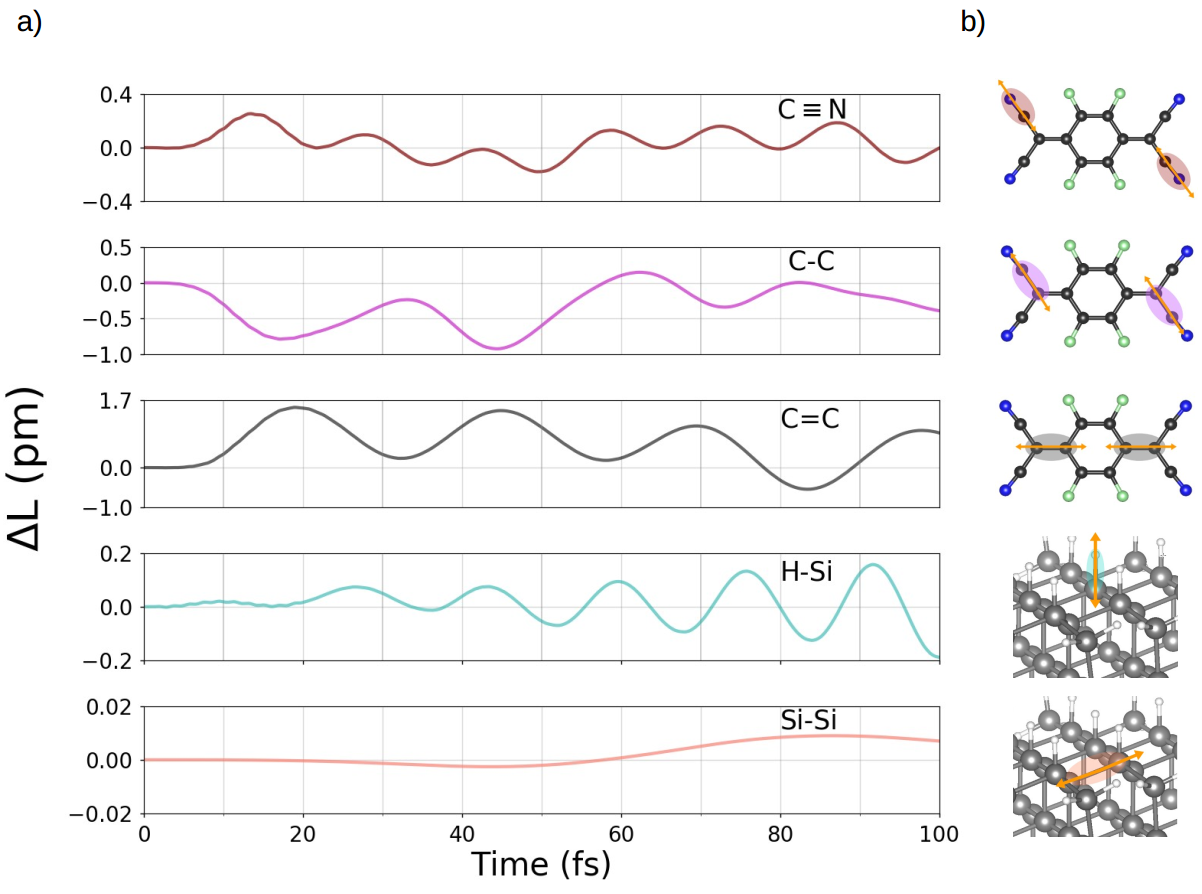}
    \caption{a) Dynamics of selected bond lengths, highlighted in panel b), computed with respect to their values in at $t=0$.}
    \label{fig:vib-dyn}
\end{figure}

Based on the results shown in Fig.~\ref{fig:MD2}b), we can filter out the contributions that are relevant in the analysis of the coupled electron-nuclear dynamics of the laser-excited hybrid interface. 
The Si-Si modes are too slow to contribute significantly in the explored time window.
his is evident also from Fig.~\ref{fig:vib-dyn}, bottom panel, where we display the variation of the Si-Si bond length compared to the ground state ($\Delta L$), as computed from RT-TDDFT coupled to the Ehrenfest dynamics.
In 100~fs the oscillation does not even complete a period and variations with respect to the ground-state bond lengths are of the order of 10$^{-4}$~\AA{}.
We can also exclude from our analysis the C-F bonds, as they do not participate in the electronic part of the excitation, which is polarized along the long molecular axis.
Hence, we focus on the dynamics of the C$\equiv$N, C--C, C=C, and H-Si stretches (see Fig.~\ref{fig:vib-dyn}).
During the first 8~fs, when the pulse is ramping up (see Fig.~\ref{fig:bader}), all bonds remain essentially unchanged with respect to their lengths at $t=0$. 
The dynamics starts after about 10~fs with all intramolecular bonds (C$\equiv$N, C--C, and C=C) oscillating almost in (anti-)phase with respect to each other until the laser is off ($t=20$~fs).
As a consequence, the dynamics of the Bader charges exhibits an oscillatory behavior in this time window, with a maximum after approximately 14~fs (see Fig.~\ref{fig:bader}b).
This initially coherent intramolecular vibrational motion begins to dephase after 20~fs, due to the higher frequency of the C$\equiv$N bond compared to those of the single and double carbon bonds. 
Concomitant with this dephasing, also the dynamics of the Bader charges becomes more incoherent (see Fig.~\ref{fig:bader}b). 
The oscillations of the C--C and C=C bonds remain approximately in anti-phase over the entire 100~fs time window.
However, the amplitude of their oscillations is different, with the double bond experiencing much larger variations compared to the single one.
We interpret this behavior as an indication of enhanced coupling of the C=C mode with the electronic part of the excitation.

Also the H-Si bonds exhibit a large number of oscillations during the explored time window, which is consistent with their high frequency seen in Fig.~\ref{fig:MD2}b). 
However, the amplitude of these oscillations is significantly smaller compared to those of the intramolecular bonds, especially in the first 50~fs, as visible from the scales of the $y$-axes in Fig.~\ref{fig:vib-dyn}a).
It should be noted that in the inorganic side of the interface the residual forces between H and Si atoms, which are present due to the relatively high threshold adopted in the structural optimization (see Section~\ref{sec:theory}), cause oscillations in the free propagation.
The latter, which are insignificant in the molecule, are subtracted from the dynamics of the Si-Si and the H-Si bonds shown in Fig.~\ref{fig:vib-dyn}a). 
The increasing amplitude of the oscillations of the H-Si bond is a manifestation of the dephased nuclear motion in the free and laser-induced dynamics.
From the results displayed in Fig.~\ref{fig:vib-dyn}, we can conclude that the vibronic coupling acts mainly on the organic side of the interface.

The scenario presented above is substantially different from previous work on all-organic interfaces, where the vibronic coupling acts coherently on either side of the interface and enhances the charge transfer from the donor to the acceptor~\cite{falk+14sci}.
Also in the F4TCNQ:H-Si(111) interface the laser-excited charge carriers tend to localize from the inorganic donor to the molecular acceptor.
However, the polarization of the transition-dipole moment parallel to the plane of the molecule additionally promotes \textit{intramolecular} charge transfer which is further enhanced by the coupling with the nuclear motion within F4TCNQ in the considered 100 fs timescale.
All in all, this effect competes against the \textit{interfacial} charge transfer and gives rise to the result shown in Fig.~\ref{fig:bader}b.

\begin{figure}[h!]
    \centering
    \includegraphics[width=1.0\linewidth]{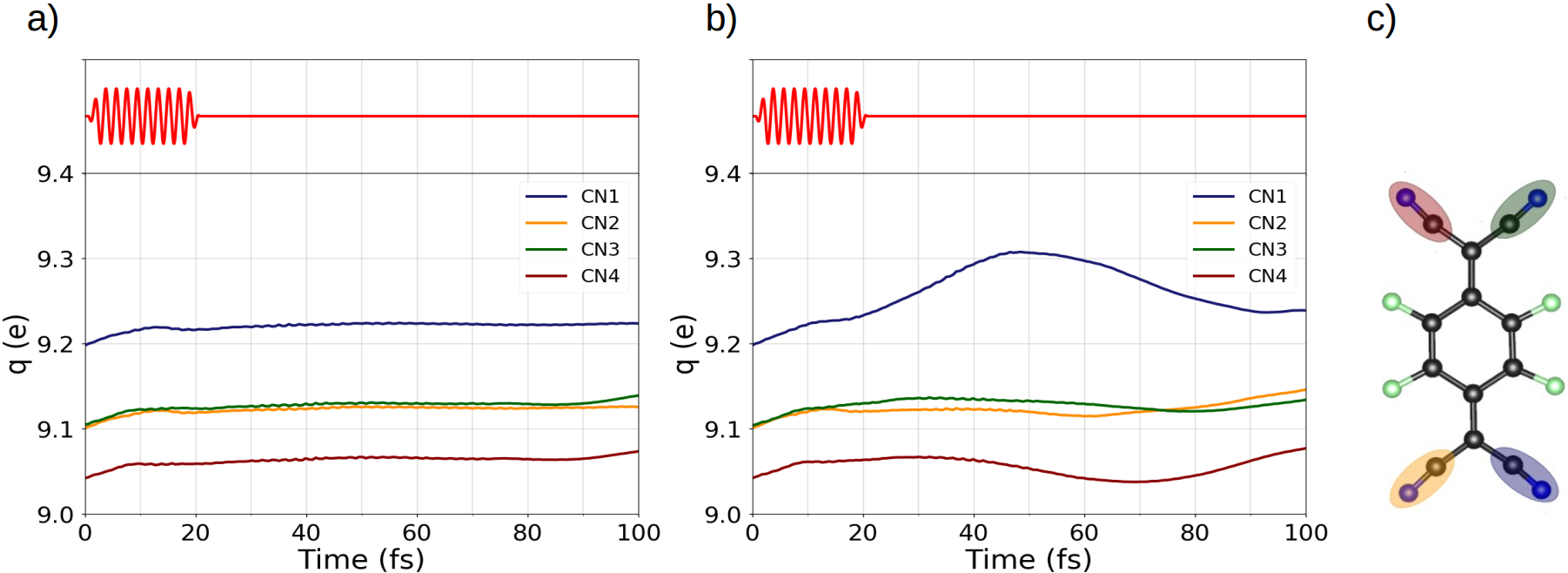}
    \caption{Time-resolved Bader charges of the C and N atoms highlighted in panel c), computed a) without and b) with the inclusion of vibronic coupling.}
    \label{fig:CN}
\end{figure}

To further investigate the physical origin of the vibronically-amplified intramolecular charge transfer, we make use again of the Bader charge analysis considering in this case a different partition related only to the organic side.
Specifically, we inspect the variation in time of the Bader charges localized on the four pairs of C and N atoms forming the triple bonds in the adsorbed F4TCNQ molecule.
Within the hybrid interface the symmetry of the molecule is broken and the C$\equiv$N bonds, which are all equivalent in the gas phase, are characterized by different lengths.
This is a signature of charge transfer in the ground state~\cite{zhu+11cm}, namely of the charge redistribution within the molecule due to its adsorption on the H-Si(111) cluster independent of the photo-excitation.
This variation of the C$\equiv$N bond length affects also the absolute values of the Bader charges at $t=0$ (see Fig.~\ref{fig:CN}a), as discussed in Section~\ref{ssec:gs}.
The C and N atoms forming bonds of equal length (CN2 and CN3 in Fig.~\ref{fig:CN}) carry the almost identical charge of 9.10~$e$ at $t=0$.
Conversely, in the other two pairs of C and N atoms (CN1 and CN4, respectively, see Fig.~\ref{fig:CN}), the Bader charges at $t=0$ amount to 9.20~$e$ and 9.05~$e$, respectively.
We note in passing that the sum of the valence electrons in the C and N atoms is equal to 9.
From a careful inspection of Fig.~\ref{fig:CN}, we notice that, when the nuclear motion is neglected (panel a), the Bader charges remain essentially constant at their ground-state values for the entire time propagation. 
On the contrary, the vibronic coupling enabled by the nuclear motion promotes charge transfer of about 0.1~$e$ compared to the ground state between the two pairs of C and N atoms at the opposite ends of F4TCNQ (see Fig.~\ref{fig:CN}b,c).
This value is of the same order of the variation of the Bader charges across the interface with respect to their ground-state value (see Fig.~\ref{fig:bader}).
The vibronic coupling induces oscillations only in the charges across the C and N atoms which exhibit different values already at $t=0$.
These oscillations are not perfectly in phase with each other, suggesting that a time window longer than 100 fs is needed to build full coherence in this process. 

\section{Summary, Conclusions, and Outlook}
In summary, we have presented a first-principles study of the laser-induced charge transfer dynamics in the hybrid interface formed by a F4TCNQ molecule physisorbed on a hydrogenated silicon cluster.
This system is $p$-doped with partially occupied HOMO and LUMO, and exhibits a charge transfer of 0.36~$e$ in the ground state, as predicted by the adopted DFT formalism.
By exciting the hybrid interface with an ultrafast laser pulse in resonance with one of its bright excitations in the visible region ($\hbar \omega = 2.18$~eV), we have followed the charge-carrier dynamics and examined the role of vibronic coupling. 
As expected, the resonant laser perturbation promotes charge transfer across the interface by about 0.1 $e$ with respect to the ground state.
However, the amount of charge transfer is not enhanced when the vibrational motion is coupled to the electron dynamics. 
The reason behind this somehow counter-intuitive behavior is related to the dominant contribution of \textit{intramolecular} vibronic coupling, which competes against the interfacial charge transfer driven by the laser-induced electronic excitation.

The scenario illustrated above is evidently dependent on the considered system, exhibiting charge transfer already in the ground state, on the photo-excitation conditions in terms of laser frequency and polarization, as well as on the temporal window of the dynamics.
However, some aspects are characteristic of hybrid interfaces and intrinsically different from all-organic complexes.
The chemical nature of the constituents, with lighter elements on the organic side and heavier atoms on the inorganic one, inhibits the coherence of the nuclear motion across the interface and hence prevents or at least slows down its coupling to the electron dynamics.
Moreover, the lowest-energy excitations in hybrid interfaces are not necessarily polarized across the interface, as it is often the case in $\pi$-$\pi$ coupled organic systems.
On the contrary, they are very much dependent on the relative orientation of the constituents, and also on their level alignment and electronic hybridization.  
For example, exciting the system in resonance with an optical transition polarized along the stacking direction, rather than parallel to it, can enhance the interfacial charge transfer.
All these variables contribute to create an intricate picture that requires an appropriate theoretical description.
The RT-TDDFT approach adopted here is capable to describe the ultrafast dynamics of hybrid interfaces and to unravel the fundamental mechanisms ruling the excitation process therein.
The explicit inclusion in the simulation of the time-dependent electric field is an added value that allows to access the transient phase of the dynamics, which is inaccessible in experiments and also in calculations on electronically constrained systems.

For all these reasons, this study and the results therein can be regarded as a starting point towards a comprehensive understanding of the charge-transfer dynamics in hybrid systems. 
RT-TDDFT has the potential to tackle this challenging task but synergistic interplay with complementary methodologies and with experiments is crucial to reach this goal. 
For example, semi-empirical approaches such as time-dependent density-functional tight binding~\cite{bona+18jpcl} can bridge the gap towards longer time scales and also enable the application of hybrid functionals for a more quantitative assessment of the charge transfer.
Moreover, methods that go beyond the mean-field Ehrenfest approach, coupling electron and nuclear dynamics~\cite{tull12jcp,agos18epjb,sato+18prb} should be employed to explore temporal ranges of the order of picoseconds and thus to assess the contribution of the inorganic side of the interface to the coupled electron-vibrational dynamics.
Additional insight can be gained by interfacing RT-TDDFT results with those obtained from model Hamiltonians, in order to include in the dynamics correlation effects, such as bound electron-hole pairs, that are not embedded in the time-dependent KS equations.
In the Si-based system considered here, excitons are not expected to play a significant role.
However, this is not the case of hybrid interfaces composed of TMDCs, where screening and excitonic effects are known to be very relevant~\cite{uged+14natm,cher+14prl,he+14prl,thyg172DM,aror+17nl,mola+19prl,lau+19prm} also in the dynamical regime~\cite{stei+15nl,seli+16natcom,sim+16natcom,pogn+16nano,moli+17nl,buad+18condmat}.
Combined efforts in the aforementioned directions will offer unprecedented insight into the fundamental mechanisms of charge-transfer dynamics in hybrid interfaces, paving the way for controlled manipulation of light-matter interaction in this class of materials on the atomistic space and time scales.

\section*{Acknowledgement}
Fruitful discussions with Carlo Andrea Rozzi are kindly acknowledged.
This work was funded by the Deutsche Forschungsgemeinschaft (DFG) - Projektnummer 182087777 - SFB 951, and 286798544 - HE 5866/2-1.
Computational resources provided by the North-German Supercomputing Alliance (HLRN), projects bep00060 and bep00076.

%

%




\end{document}